\documentclass[preprint,preprintnumbers,amsmath,amssymb]{revtex4}

\usepackage{dcolumn}
\usepackage{bm}

\begin{document}

\title{An understanding for flavor physics in the lepton sector}

\author{Zhen-hua Zhao}
\affiliation{
Institute of Theoretical Physics, Chinese Academy of Sciences, \\
and State Key Laboratory of Theoretical Physics,\\
P. O. Box 2735, Beijing 100190, China}
 \email{zhzhao@itp.ac.cn}

\date{\today}

\begin{abstract}

In this paper, we give a model for understanding flavor physics in the lepton sector--mass hierarchy among different generations and neutrino mixing pattern.
The model is constructed in the framework of supersymmetry, with a family symmetry $S4*U(1)$.
There are two right-handed neutrinos introduced for seesaw mechanism, while some standard model(SM) gauge group singlet fields are included which transforms non-trivially under family symmetry.
In the model, each order of contributions are suppressed by $\delta \sim 0.1$ compared to the previous one.
In order to reproduce the mass hierarchy, $m_\tau$ and $\sqrt{\Delta m_{atm}^2}$, $m_\mu$ and $\sqrt{\Delta m_{sol}^2}$ are obtained at leading-order(LO) and next-to-leading-order(NLO) respectively, while electron can only get its mass through next-to-next-to-next-to-leading-order(NNNLO) contributions.
For neutrino mixing angels, $\theta_{12}, \theta_{23}, \theta_{13}$ are $45^\circ, 45^\circ, 0$ i.e. Bi-maximal mixing pattern as first approximation, while higher order contributions can make them consistent with experimental results.
As corrections for $\theta_{12}$ and $\theta_{13}$ originate from the same contribution, there is a relation predicted for them $\sin{\theta_{13}}=\displaystyle \frac{1-\tan{\theta_{12}}}{1+\tan{\theta_{12}}}$.
Besides, deviation from $\displaystyle \frac{\pi}{4}$ for $\theta_{23}$ should have been as large as deviation from 0 for $\theta_{13}$ if it were not the former is suppressed by a factor 4 compared to the latter.

\end{abstract}


\maketitle

\setlength{\parskip}{0.5\baselineskip}
\section{introduction}

Before, many believed that $\theta_{13}$ is very small due to the fact that experiments could only give an upper bound for it for a long time.
In this situation, the scenario with the so-called Tri-bimaximal mixing pattern \cite{tri-bi} which gives a vanishing $\theta_{13}$ as LO approximation was very popular.
Remarkably, it is found that many models with a discrete flavor symmetry can realize this mixing pattern from an underlying theory \cite{ma,tri bi}.
However, considering latest experimental results for $\theta_{13}$ \cite{daya-bay,reno,t2k,minos,chooz} which turned out to be much larger than many expected, Tri-bimaximal model needs substantial modification.
The main problem with it lies in that it is difficult to understand how and why  NLO corrections take $\theta_{13}$ from 0 to a rather large angle while preserving $\theta_{12}$ and $\theta_{23}$ close to their values given at LO.

In this paper, we would like to give a simple alternative to Tri-bimaximal model.
For convenience of expressing our ansatz for leptons' mass matrices, a model realizing our ansatz is given.
In the following, this realistic model is given directly, while observations about lepton flavor physics are given implicitly under discussions for the model.
In the model, mass hierarchy in the lepton sector as well as neutrino mixing pattern are natural results.
To reproduce the mass hierarchy, fermion masses are produced in different orders of contributions.
Because of special forms for mass matrices, realistic neutrino mixing pattern will also be obtained.

\section{the model}
The model is built in the framework of supersymmetry with family symmetry $S4*U(1)$; field contents and their transformation properties under $S4*U(1)*U(1)_R$ are given in Table 1.
$S4$ group has 2 singlet representations, 1 doublet representations and 2 triplet representations which are denoted as $1$, $1^\prime$, $2$, $3$ and $3^\prime$ in order.
For its presentation, we will adopt the presentation reported in appendix of \cite{model1} where readers can find details about multiplication rules and Clebsch-Gordan coefficients.
(For a recent review about models for flavor physics with $S4$ family symmetry, please see \cite{s4}.)
As shown in Table 1, we require that three $SU(2)_L$ doublets combine to be representation 3 under $S4$, $e^c$ and ($\mu^c$,$\tau^c$) transform as representation $1$ and $2$ respectively, while right handed neutrino $N_1$ and $N_2$ are in representation $1$ and $1^\prime$.
Higgs fields $H_{u,d}$ are trivial representations under family symmetry, while some SM singlet flavon fields which fall in non-trivial representations of $S4$ are included.

\begin{table}[tbp]
\centering
\begin{tabular}{lcccccccccccccc}
\hline
         &$L$ &$e^c$ &($\mu^c$,$\tau^c$)&$N_1$ &$N_2$            &$H_{u,d}$ &$\Psi$     &$\Omega$   &$\Sigma$ &$\Phi$      &$\Pi^0$  &$\Theta^0$ \\ \hline
$S4$     &3   &1     &2                 &1     &$1^\prime$       &1         &$3^\prime$ &$3^\prime$ &2        &$3^\prime$  &2        &3           \\
$U(1)$    &2  &16    &3                 &0     &0                &0         &-1         &-2          &-3       &-5          &3        &10           \\
$U(1)_R$ &0   &1     & 1                &1     &1                &1         &0          &0          &0        &0           &2        &2            \\ \hline
\end{tabular}
\caption{Transformation properties of all the fields under $S4*U(1)*U(1)_R$.}
\end{table}

Due to the transformation properties distributed above, LO terms that can generate masses for leptons are of dimension 5.
Besides, only when flavon fields get vacuum expectation values(VEVs) can fermion masses be produced.
In this situation, terms contributing to fermion masses are characterized by $\displaystyle (\frac{v}{M})^n$ where $v$ denotes flavon fields' VEVs, $M$ is cutoff scale for flavon physics and $n$ indicates number of flavon fields.
In following discussions, we assume that $\delta=\displaystyle \frac{v}{M} \sim 0.1$.
As a result, the order a term belongs to can be classified by the number of flavon fields.

A $U(1)$ symmetry is also introduced which plays a similar role as Froggate-Nelson symmetry \cite{f-n}.
With family symmetry $U(1)$, appropriate flavon fields can be picked out for different leptons to produce mass matrices that is needed.
Besides, R-symmetry is included which plays a key part in discussing flavon fields' VEVs.
As flavon fields have 0 charge under R-symmetry, they have to appear in companion with a driving field which is marked with a suffix $0$ in Table 1.
Consequently, supersymmetric condition that $F$ components of driving fields cannot have VEVs provides constrains on flavon fields' VEVs.

For the time being, we just assume flavon fields' VEVs have the following form and are stable against higher order contributions,
\begin{equation}
    \langle\Psi\rangle=\left(\begin{array}{c}
     v_1 \\
      2v_1\\
     0
\end{array}
\right)\hspace{0.5cm}
\langle\Omega\rangle=\left(\begin{array}{c}
     0 \\
      v_2\\
    0
\end{array}
\right)\hspace{0.5cm}
\langle\Sigma\rangle=\left(\begin{array}{c}
      0\\
      v_3
\end{array}
\right)\hspace{0.5cm}
\langle\Phi\rangle=\left(\begin{array}{c}
     0\\
     v_4\\
     v_4
\end{array}
\right),
\end{equation}
where all the VEVs are assumed to be close to each other $v_1 \sim v_2 \sim v_3 \sim v_4 \sim v$.
In the end of this section, we will justify these VEV alignments.

\subsection{Physics at Leading Order}
At LO, superpotential includes the following terms that contribute to lepton masses,
\begin{equation}
\displaystyle \frac{1}{\Lambda} y_1 [(\mu^c,\tau^c)L]_{3^\prime} \Phi H_d + \displaystyle \frac{1}{\Lambda} y_2 N_2 L \Omega H_u + M_1 N_1 N_1+M_2 N_2 N_2,
\end{equation}
where we use $[\ \ \ ]_{3^\prime}$ to indicate that $(\mu^c,\tau^c)L$ combines to become representation $3^\prime$ and so on.
In this work, all dimensionless coupling such as $y_1$ and $y_2$ in Eq.(2) are assumed to be order 1 and close to each other.
With VEVs in Eq.(1), mass matrices for charged leptons and light neutrinos are as follows,
\begin{equation}
M_e=\displaystyle \frac{y_1 v_d v_4}{\Lambda}\left(\begin{array}{ccc}
     0& 0 & 0\\
     0& \displaystyle \frac{\sqrt 3}{2} &\displaystyle \frac{\sqrt 3}{2}\\
      0&   \displaystyle \frac{1}{2}  & \displaystyle \frac{1}{2}
\end{array}
\right), \hspace{1cm}
M_\nu=\displaystyle \frac{(y_1 v_u v_2)^2}{M_2\Lambda^2}\left(\begin{array}{ccc}
     0& 0 & 0\\
     0& 0 &0\\
      0& 0 & 1
\end{array}
\right).
\end{equation}
$M_e^\dag M_e$ have the following form,
\begin{equation}
M_e^\dag M_e=\displaystyle (\frac{y_1 v_d v_4}{\Lambda})^2\left(\begin{array}{ccc}
     0& 0 & 0\\
     0& 1 &1\\
      0& 1 & 1
\end{array}
\right).
\end{equation}
Thus, only tau and one neutrino have non-zero masses $m_\tau=\sqrt2 \displaystyle \frac{y_1 v_d v_4}{\Lambda}$ and $m_3=\displaystyle \frac{(y_1 v_u v_2)^2}{M_2\Lambda^2}$.

\subsection{Physics at Next to Leading Order}
After taking NLO contributions into consideration, there are new terms contributing to lepton masses,
\begin{equation}
\displaystyle \frac{1}{\Lambda^2} y_3 [(\mu^c,\tau^c)L]_3 [\Omega \Sigma]_3 H_d + \displaystyle \frac{1}{\Lambda^2} y_4 [(\mu^c,\tau^c)L]_{3^\prime} [\Omega \Sigma]_{3^\prime} H_d + \displaystyle \frac{1}{\Lambda^2} y_5 N_1 L [\Psi \Psi]_3 H_u.
\end{equation}
As a result, charged lepton and light neutrino mass matrices become,
\begin{equation}
M_e^\prime=\displaystyle \frac{y_1 v_d v_4}{\Lambda}\left(\begin{array}{ccc}
     0& 0 & 0\\
     0& \displaystyle \frac{\sqrt 3}{2} &\displaystyle \frac{\sqrt 3}{2}-\displaystyle \frac{\delta_1}{4} +\displaystyle \frac{3\delta_2}{4} \\
      0&   \displaystyle \frac{1}{2}+\displaystyle \frac{\sqrt3 \delta_1}{4}+ \displaystyle \frac{\sqrt3 \delta_2}{4} & \displaystyle \frac{1}{2}
\end{array}
\right),
\end{equation}
where $\delta_1= \displaystyle \frac{y_3 v_2 v_3}{y_1 v_4 \Lambda} \sim \delta_2= \displaystyle \frac{y_4 v_2 v_3}{y_1 v_4 \Lambda} \sim \delta$;
\begin{equation}
M_\nu^\prime=\displaystyle \frac{(y_1 v_u v_2)^2}{M_2\Lambda^2}\left(\begin{array}{ccc}
     16 \delta_3^2& 16 \delta_3^2 & 0\\
     16 \delta_3^2& 16 \delta_3^2 &0\\
      0& 0 & 1
\end{array}
\right),
\end{equation}
where $\delta_3^2=\displaystyle \frac{M_2}{M_1}\displaystyle (\frac{y_5 v_1 v_1}{y_2 v_2 \Lambda})^2 \sim \delta^2$.

There will be no higher orders of contributions to neutrino mass, so Eq.(7) is the final result.
The following matrix diagonalize Eq.(7) to obtain three mass eigenvalues $m_1=0, m_2= 32 \delta_3^2 \displaystyle \frac{(y_1 v_u v_2)^2}{M_2\Lambda^2}, m_3=\displaystyle \frac{(y_1 v_u v_2)^2}{M_2\Lambda^2}$,
\begin{equation}
U_\nu^\prime=\left(\begin{array}{ccc}
     \displaystyle \frac{1}{\sqrt{2}} & \displaystyle \frac{1}{\sqrt{2}}&0\\
      -\displaystyle \frac{1}{\sqrt{2}}& \displaystyle \frac{1}{\sqrt{2}}&0\\
      0&0&1
\end{array}
\right).
\end{equation}
With experimental results for neutrino oscillations \cite{global}, $\displaystyle \frac{m_2}{m_3}= 32 \delta_3^2 =\displaystyle \sqrt{\frac{\Delta m_{sol}^2}{\Delta m_{atm}^2}}= 0.18$.
Noteworthy, in mass matrix for light neutrinos, NLO contributions should have been 2 orders smaller than LO ones after seesaw mechanism.
The factor 16 arising from C-G coefficients in Eq.(7) plays a crucial role in making $\displaystyle \frac{m_2}{m_3}$ consistent with experimental result without fine-tuning.

Intuitively, smaller eigenvalue of $M_e^{\prime\dag} M_e^\prime$ should be about $\delta \displaystyle (\frac{y_1 v_d v_4}{\Lambda})^2$.
However, it can be proved that smaller eigenvalue of $M_e^{\prime\dag} M_e^\prime$ is about $\delta^2 \displaystyle (\frac{y_1 v_d v_4}{\Lambda})^2$ while the larger one remains about $2 \displaystyle (\frac{y_1 v_d v_4}{\Lambda})^2$.
Thus, $\displaystyle \frac{m_\mu}{m_\tau} \approx \displaystyle \sqrt{\frac{\delta^2}{2}}$, compatible with experimental result 0.06.
As for $U_e^\prime$, only $\theta_{23}$ is non-zero and there is an estimate for it,
\begin{equation}
\tan{2\theta_{23}} \approx \displaystyle \frac{2}{\delta}.
\end{equation}
If we parameterize deviation from $\displaystyle \frac{\pi}{4}$ for $\theta_{23}$ with a small quantity $\epsilon_1$, it is about $\displaystyle \frac{\delta}{4}$ and $U_e^\prime$ can be described in the following form,
\begin{equation}
U_e^\prime=\left(\begin{array}{ccc}
     1& 0 & 0\\
     0& \displaystyle \frac{1}{\sqrt{2}}(1+\epsilon_1)& \displaystyle \frac{1}{\sqrt{2}}(1-\epsilon_1)\\
      0& -\displaystyle \frac{1}{\sqrt{2}}(1-\epsilon_1)& \displaystyle \frac{1}{\sqrt{2}}(1+\epsilon_1)
\end{array}
\right).
\end{equation}

\subsection{Physics at Next-to-Next-to-Leading-Order}

New terms that contribute to lepton masses are listed below,
\begin{equation}
\begin{split}
\displaystyle \frac{1}{\Lambda^3} y_6 [(\mu^c,\tau^c)L]_3 [(\Omega \Omega)_2 \Psi]_3 H_d+\displaystyle \frac{1}{\Lambda^3} y_7 [(\mu^c,\tau^c)L]_3 [(\Omega \Omega)_3 \Psi]_3 H_d \\+\displaystyle \frac{1}{\Lambda^3} y_8 [(\mu^c,\tau^c)L]_{3^\prime} [(\Omega \Omega)_2 \Psi]_{3^\prime} H_d +\displaystyle \frac{1}{\Lambda^3} y_9 [(\mu^c,\tau^c)L]_3 [\Sigma (\Psi \Psi)_3]_3 H_d.
\end{split}
\end{equation}
At this stage, mass matrix for charged leptons becomes,
\begin{equation}
M_e^{\prime\prime}=\displaystyle \frac{y_1 v_d v_4}{\Lambda}\left(\begin{array}{ccc}
     0& 0 & 0\\
     -\sqrt 3\delta_4^2& \displaystyle \frac{\sqrt 3}{2} &\displaystyle \frac{\sqrt 3}{2}-\displaystyle \frac{\delta_1}{4} +\displaystyle \frac{3\delta_2}{4} \\
      0&   \displaystyle \frac{1}{2}+\displaystyle \frac{\sqrt3 \delta_1}{4}+ \displaystyle \frac{\sqrt3 \delta_2}{4} & \displaystyle \frac{1}{2}
\end{array}
\right),
\end{equation}
where $\delta_4^2= \displaystyle \frac{y_6 v_1 v_2 v_2}{y_1 v_4 \Lambda \Lambda} \sim \delta^2$.
In Eq.(12), we have neglected contributions to those matrix elements which are non-zero at NLO.

In order to make physics clear, we do a qualitative analysis for $U_e^{\prime \prime}$ up to matrix elements' orders.
First of all, we effect transformation $U_e^\prime$ on $M_e^{\prime \prime \dag} M_e^{\prime \prime}$,
\begin{equation}
U_e^{\prime \dag}M_e^{\prime \prime \dag} M_e^{\prime \prime}U_e^\prime \approx \displaystyle \frac{y_1 v_d v_4}{\Lambda}\left(\begin{array}{ccc}
     \delta^4& \delta^3& \delta^2\\
     \delta^3 & \delta^2 &0 \\
      \delta^2&  0 & 2
\end{array}
\right).
\end{equation}
To diagonalize Eq.(13), we just need to effect a transformation with $\sin{\theta_{13}} \sim \delta^2$ and a transformation with $\sin{\theta_{12}} \sim \delta$ successively.
If we ignore the negligibly small $\theta_{13}$ and parameterize $\theta_{12}$ with another small quantity $\epsilon_2$, $U_e^{\prime \prime}$ have the following form,
\begin{equation}
U_e^{\prime \prime}=\left(\begin{array}{ccc}
     1& \epsilon_2 & 0\\
     -\displaystyle \frac{\epsilon_2}{\sqrt{2}}& \displaystyle \frac{1}{\sqrt{2}}(1+\epsilon_1)& \displaystyle \frac{1}{\sqrt{2}}(1-\epsilon_1)\\
      \displaystyle \frac{\epsilon_2}{\sqrt{2}}& -\displaystyle \frac{1}{\sqrt{2}}(1-\epsilon_1)& \displaystyle \frac{1}{\sqrt{2}}(1+\epsilon_1)
\end{array}
\right).
\end{equation}

At present, we can discuss about neutrino mixing angles.
$U_{PMNS}$ \cite{pmns} is obtained by $U_e^{\prime \prime \dag} U_\nu^\prime$,
\begin{equation}
U=\left(\begin{array}{ccc}
     \displaystyle \frac{1}{\sqrt{2}}(1+\displaystyle \frac{1}{\sqrt{2}}\epsilon_2)&\displaystyle \frac{1}{\sqrt{2}}(1-\displaystyle \frac{1}{\sqrt{2}}\epsilon_2)& \displaystyle \frac{\epsilon_2}{\sqrt{2}} \\
     \cdots &\cdots & \displaystyle \frac{1}{\sqrt{2}}(1-\epsilon_1)\\
     \cdots& \cdots& \displaystyle \frac{1}{\sqrt{2}}(1+\epsilon_1)
\end{array}
\right).
\end{equation}
where we just list the matrix elements involved in fixing neutrino mixing angles.
In this case, $\tan{\theta_{12}}=\displaystyle \frac{1-\displaystyle \frac{1}{\sqrt{2}}\epsilon_2}{1+\displaystyle \frac{1}{\sqrt{2}}\epsilon_2},  \sin{\theta_{13}}=\displaystyle \frac{\epsilon_2}{\sqrt{2}}, \tan{\theta_{23}}=\displaystyle \frac{1-\epsilon_1}{1+\epsilon_1}$.
The larger the deviation from $\displaystyle \frac{\pi}{4}$ for $\theta_{12}$ the larger $\theta_{13}$ is, resulting from the relation $\sin{\theta_{13}}=\displaystyle \frac{1-\tan{\theta_{12}}}{1+\tan{\theta_{12}}}$.
With result for $\theta_{12}$ \cite{global}, $\sin{\theta_{13}}$ is obtained as $0.188^{+0.015}_{-0.017}$, which is a little larger than Daya Bay's result $0.153^{+0.018}_{-0.019}$ and consistent with RENO's result $0.171^{+0.023}_{-0.027}$ and Double Chooz's result $0.167^{+0.040}_{-0.050}$.
Noteworthy, order $\delta^2$ contribution which is neglected in obtaining Eq.(14) can lead to additional $O(0.01)$ contribution to $\sin{\theta_{13}}$, making the model consistent with experimental results.
As far as $\theta_{23}$ is concerned, its deviation from $\displaystyle \frac{\pi}{4}$ should have been as large as $\theta_{12}$'s deviation from $\displaystyle \frac{\pi}{4}$ and $\theta_{13}$'s deviation from 0, if it were not for the fact the former is suppressed by additional factor 4.
Therefore, $\theta_{23}$'s deviation from $\displaystyle \frac{\pi}{4}$ is a little smaller than $\theta_{12}$'s deviation from $\displaystyle \frac{\pi}{4}$ but is too large for $\theta_{23}$ to be taken as maximal.

\subsection{Physics at Next-to-Next-to-Next-to-Leading-Order}

Electron cannot obtain its mass until this order where interacting terms between $e^c$ and $L$ appears for the first time,
\begin{equation}
\begin{split}
\displaystyle \frac{1}{\Lambda^4}[e^c L]_3 H_d\{y_{10}[(\Phi \Phi)_1  (\Sigma \Phi)_3]_3+y_{11} [(\Phi \Phi)_2 (\Sigma \Phi)_3]_3+y_{12} [(\Phi \Phi)_2  (\Sigma \Phi)_{3^\prime}]_3\\+y_{13}[(\Phi \Phi)_3  (\Sigma \Phi)_3]_3 +y_{14}[(\Phi \Phi)_3  (\Sigma \Phi)_3]_{3^\prime} + y_{15} [(\Phi \Phi)_{3^\prime}  (\Sigma \Phi)_3]_3
+y_{16} [(\Phi \Phi)_{3^\prime}  (\Sigma \Phi)_{3^\prime}]_{3^\prime}\}.
\end{split}
\end{equation}
Charged leptons' mass matrix now becomes,
\begin{equation}
M_e^{\prime\prime\prime}=\displaystyle \frac{y_1 v_d v_4}{\Lambda}\left(\begin{array}{ccc}
     0& \delta_5^3+\delta_6^3 & \delta_5^3+\delta_6^3 \\
     -\sqrt 3\delta_4^2& \displaystyle \frac{\sqrt 3}{2} &\displaystyle \frac{\sqrt 3}{2}-\displaystyle \frac{\delta_1}{4} +\displaystyle \frac{3\delta_2}{4} \\
      0&   \displaystyle \frac{1}{2}+\displaystyle \frac{\sqrt3 \delta_1}{4}+ \displaystyle \frac{\sqrt3 \delta_2}{4} & \displaystyle \frac{1}{2}
\end{array}
\right),
\end{equation}
where $\delta_5^3= \displaystyle \frac{y_{10} v_3 v_4 v_4}{y_1  \Lambda \Lambda \Lambda} \sim \delta_6^3=\displaystyle \frac{y_{11} v_3 v_4 v_4}{y_1 \Lambda \Lambda \Lambda} \sim \delta^3$.
Eq.(17) leads to a naturally small electron mass, without interfering the above discussions.

\subsection{Flavon Fields' VEVs}

Finally, we would like to address issues concerning VEV alignments in Eq.(1) which needs to be a reasonable result in order to make this model convincing.
We just need to show that these VEV alignments are valid up to NNLO, because our physical results except electron mass have been achieved by this order while production of electron mass does not rely on those specific VEV alignments.
As we have said, supersymmetric requirements result in constraint on F components of driving fields $\Pi^0$ and $\Theta^0$ that $\langle F_i\rangle=0$ with $i$ representing $\Pi^0_1,\Pi^0_2,\Theta^0_1,\Theta^0_2,\Theta^0_3$.
Relevant terms are given below:\\
LO,\ \ \ \ \ $\Pi^0\{m\Sigma+c_1(\Omega\Psi)_2 \}+\Theta^0\{c_4(\Phi\Phi)_3\}$;\\
NLO,\ \ \ \ \ \ $\displaystyle \frac {1}{\Lambda} \Pi^0 \{c_2 [(\Psi\Psi)_3\Psi]_2+c_3 [(\Psi\Psi)_{3^\prime}\Psi]_2\}+\displaystyle \frac {1}{\Lambda} \Theta^0\{c_5 [(\Sigma\Phi)_3\Omega]_3+c_6 [(\Sigma\Phi)_{3^\prime}\Omega]_3\}$;\\
NNLO, \ \ \ \ \ $\displaystyle \frac {1}{\Lambda^2} \Theta^0 \{c_7 \{[(\Sigma\Phi)_{3}\Psi]_{1^\prime}\Psi\}_{3}+c_8 \{[(\Sigma\Phi)_{3}\Psi]_{2}\Psi\}_{3}+c_9 \{[(\Sigma\Phi)_{3}\Psi]_{3}\Psi\}_{3}\\
+c_{10} \{[(\Sigma\Phi)_{3}\Psi]_{3^\prime}\Psi\}_{3}+c_{11} \{[(\Sigma\Phi)_{3^\prime}\Psi]_{2}\Psi\}_{3}+c_{12} \{[(\Sigma\Phi)_{3^\prime}\Psi]_{3}\Psi\}_{3}+c_{13} \{[(\Sigma\Phi)_{3^\prime}\Psi]_{3^\prime}\Psi\}_{3}\\
c_{14} \{[(\Omega\Omega)_{1}\Phi]_{3^\prime}\Psi\}_{3}+c_{15} \{[(\Omega\Omega)_{2}\Phi]_{3}\Psi\}_{3}+c_{16} \{[(\Omega\Omega)_{2}\Phi]_{3^\prime}\Psi\}_{3}+c_{17} \{[(\Omega\Omega)_{3}\Phi]_{1^\prime}\Psi\}_{3}\\
+c_{18} \{[(\Omega\Omega)_{3}\Phi]_{2}\Psi\}_{3}+c_{19} \{[(\Omega\Omega)_{3}\Phi]_{3}\Psi\}_{3}+c_{20} \{[(\Omega\Omega)_{3}\Phi]_{3^\prime}\Psi\}_{3}+c_{21} \{[(\Omega\Omega)_{3^\prime}\Phi]_{2}\Psi\}_{3}\\
+c_{22} \{[(\Omega\Omega)_{3^\prime}\Phi]_{3}\Psi\}_{3}+c_{23} \{[(\Omega\Omega)_{3^\prime}\Phi]_{3^\prime}\Psi\}_{3}+c_{24} \{[(\Sigma\Sigma)_{1}\Sigma]_{2}\Psi\}_{3}+c_{25} \{[(\Sigma\Sigma)_{1^\prime}\Sigma]_{2}\Psi\}_{3}\\
+c_{26} \{[(\Sigma\Sigma)_{2}\Sigma]_{1^\prime}\Psi\}_{3}+c_{27} \{[(\Sigma\Sigma)_{2}\Sigma]_{2}\Psi\}_{3}
+c_{28} \{[(\Sigma\Sigma)_{1}\Omega]_{3^\prime}\Omega\}_{3}+c_{29} \{[(\Sigma\Sigma)_{1^\prime}\Omega]_{3}\Omega\}_{3}+c_{30} \{[(\Sigma\Sigma)_{2}\Omega]_{3}\Omega\}_{3}+c_{31} \{[(\Sigma\Sigma)_{2}\Omega]_{3^\prime}\Omega\}_{3}$.

There are totally 13 equations for $\langle F_i\rangle=0$ at LO, NLO, NNLO respectively, among which 8 equations are automatically satisfied when flavon fields take VEVs as shown in Eq.(1).
The other 5 equations are as follows,
\begin{eqnarray}
\left\{
\begin{array}{lll}
m v_3+ \sqrt 3 c_1 v_1 v_2&=& 0 \\
 (c_5-\sqrt 3 c_6)v_2 v_3 v_4&=& 0\\
\lambda_1 v_1 v_1 v_3 v_4+\lambda_2 v_1 v_2 v_2 v_4+\lambda_3 v_1 v_3 v_3 v_3+\lambda_4 v_2 v_2 v_3 v_3&=&0\\
\lambda_5 v_1 v_1 v_3 v_4+\lambda_6 v_1 v_2 v_2 v_4+\lambda_7 v_1 v_3 v_3 v_3&=& 0\\
\lambda_8 v_1 v_1 v_3 v_4+\lambda_9 v_1 v_2 v_2 v_4&=& 0
\end{array},
\right.
\end{eqnarray}
where $\lambda_1=c_7-\displaystyle \frac{1}{2}c_8-c_9-c_{10}-\displaystyle \frac{\sqrt3}{2}c_{11}-\sqrt3 c_{12}-\sqrt3 c_{13},\\
\lambda_2=\sqrt3 c_{15}-3 c_{16}-2c_{19}+2c_{20},\\
\lambda_3=-c_{24}+c_{26}-c_{27},\\
\lambda_4=-c_{28}+\displaystyle \frac{\sqrt3}{2} c_{30}+\displaystyle \frac{1}{2} c_{31},\\
\lambda_5=2 c_7+\displaystyle \frac{1}{2}c_8-\displaystyle \frac{3}{2}c_9-\displaystyle \frac{1}{2} c_{10}+\displaystyle \frac{\sqrt3}{2}c_{11}+\displaystyle \frac{5\sqrt3}{2} c_{12}-\displaystyle \frac{\sqrt3}{2} c_{13},\\
\lambda_6=\displaystyle \frac{\sqrt3}{2} c_{15}+\displaystyle \frac{3}{2} c_{16}+c_{19}+c_{20},\\
\lambda_7=c_{24}+2c_{26}+c_{27},\\
\lambda_8=\displaystyle \frac{3}{2}c_8+\displaystyle \frac{1}{2}c_9+\displaystyle \frac{3}{2} c_{10}-\displaystyle \frac{3}{2}c_{11}+\displaystyle \frac{\sqrt3}{2} c_{12}+\displaystyle \frac{3\sqrt3}{2} c_{13},\\
\lambda_9=-\displaystyle \frac{\sqrt3}{2} c_{15}-\displaystyle \frac{3}{2} c_{16}+c_{19}+c_{20}.$\\
For the second equation in Eq.(18), we have to assume an accidental relation $c_5-\sqrt 3 c_6=0$.
In this situation, values of $v_1-v_4$ are determined from the other 4 equations and should be in the order of $m$ which is the only coefficient that has dimension.
Thus, we can say the assumption $v_1\sim v_2 \sim v_3 \sim v_4 \sim v$ is reasonable.

\section{conclusions and discussions}

In conclusion, we have built a model for understanding flavor physics in the lepton sector, mainly mass spectrum and mixing pattern.
The model is constructed under family symmetry $S4*U(1)$.
With the assumption that higher order contribution is suppressed by $\delta\sim0.1$  compared to previous one, the mass hierarchy $\displaystyle \frac{m_e}{m_\mu}, \displaystyle \frac{m_\mu}{m_\tau}, \displaystyle \sqrt{\frac{\Delta m_{sol}^2}{\Delta m_{atm}^2}}$ are natural results of this model.
This is realized by producing $m_\tau, m_3$ at LO, $m_\mu, m_2$ at NLO and $m_e$ at NNNLO while $m_1=0$ to all orders.
In fact, this realization is to some extent inspired by Chun Liu's works \cite{cl} where electron, muon and tau get their mass from breaking of different symmetries.
At the same process of reproducing mass spectrum, realistic mixing pattern is obtained(in Refs.\cite{zzx}, the authors also attempted to connect mixing angles with mass hierarchy).
As a matter of fact, this model's mixing pattern for first approximation is actually Bi-maximal \cite{bi-maximal,model2}.
There are some works \cite{fritzsch,merlo,datta}obtaining Bi-maximal mixing pattern with the same starting point as this model.
(After finishing this work, we have received some works \cite{ge} which are related to ours.)
As far as mixing angels are concerned, this work does not only provide a realistic model where higher order contributions change $\theta_{13}$ and $\theta_{12}$ considerably without interfering $\theta_{23}$ much, but also predicts a relation  $\sin{\theta_{13}}=\displaystyle \frac{1-\tan{\theta_{12}}}{1+\tan{\theta_{12}}}$.
However, there is still one problem in this model i.e. the unnatural relation $c_5-\sqrt 3 c_6=0$.
Nevertheless, this problem is dependent on our choice of family symmetry and field contents.
It is possible another model where field contents or even family symmetry are different from here can realize our ansatz for leptons' mass matrices as described in this paper, without getting in trouble with unnatural relation like here when discussing flavon fields' VEVs.

\begin{acknowledgments}
I would like to thank Chun Liu for helpful discussions. This work was supported in part by the National Natural Science Foundation of China under Nos. 11075193 and 10821504, and by the National Basic Research Program of China under Grant No. 2010CB833000.
\end{acknowledgments}

\end{document}